%% file: main.tex
\documentclass[fleqn,usenatbib]{mnras}
\pdfoutput=1
\usepackage{newtxtext,newtxmath}
\usepackage{xcolor}
\usepackage{multirow}
\usepackage{threeparttable}
\usepackage{subcaption}
\usepackage{mwe}
\usepackage{comment}

\usepackage[T1]{fontenc}

\DeclareRobustCommand{\VAN}[3]{#2}
\let\VANthebibliography\thebibliography
\def\thebibliography{\DeclareRobustCommand{\VAN}[3]{##3}\VANthebibliography}

\usepackage{graphicx,caption}	
\usepackage{amsmath}	

\usepackage{footnote}
\makesavenoteenv{table}
\makesavenoteenv{tabular}
\usepackage{graphicx}
\usepackage{soul}
\usepackage{hyperref}

 \newcommand{\src}{PSR J1820--0427 }

\title[Single pulse study of PSR J1820--0427]{Single-pulse analysis and average emission characteristics of PSR J1820--0427 from observations made with the MWA and uGMRT}

\author[P. Janagal et al.]{Parul Janagal,$^{1}$\thanks{E-mail: phd1801121004@iiti.ac.in}
Manoneeta Chakraborty,$^{1}$
N. D. Ramesh Bhat,$^{2}$ 
Samuel J. McSweeney$^{2}$
\newauthor
Susmita Sett$^{2}$
\\
$^{1}$Department of Astronomy, Astrophysics, and Space Engineering, Indian Institute of Technology Indore, Indore 453552, India\\
$^{2}$International Centre for Radio Astronomy Research, Curtin University, Bentley, WA 6102, Australia\\
}

\date{Accepted XXX. Received YYY; in original form ZZZ}

\pubyear{2023}

\begin{document}
\label{firstpage}
\pagerange{\pageref{firstpage}--\pageref{lastpage}}
\maketitle

\input{Abstract}




\input{Introduction}
\input{Observations}
\input{DataProcessing}

\input{Analysis.tex}
\input{Discussion}

\input{Conclusion}

\section*{Acknowledgements}

PJ acknowledges the Senior Research Fellowship awarded by the Council of Scientific \& Industrial Research, India. We thank the staff of the GMRT and MWA who have made these observations possible. The GMRT is run by the National Centre for Radio Astrophysics of the Tata Institute of Fundamental Research. We would like to thank Bhaswati Bhattacharya for conducting the uGMRT observations. The scientific work made use of Inyarrimanha Ilgari Bundara, the CSIRO Murchison Radio-astronomy Observatory. We acknowledge the Wajarri Yamaji people as the traditional owners of the Observatory site. This work was supported by resources provided by the Pawsey Supercomputing Centre with funding from the Australian Government and the Government of Western Australia. We thank Marcin Sokolowski for his detailed comments. PJ acknowledges Akriti Sinha for her help in uGMRT image processing.

\section*{Data Availability}

This paper includes data taken from the uGMRT in the 38th observing cycle and the data taken from MWA under the project G0071. 
 



\bibliographystyle{mnras}
\bibliography{references} 








\bsp	
\label{lastpage}
\end{document}

%% file: Abstract.tex
\begin{abstract}
We have studied the pulse-to-pulse variability in PSR J1820--0427 and its frequency dependence using high-quality, wide-band observations made from the upgraded Giant Metrewave Radio Telescope (uGMRT; 300-750 MHz) and the Murchison Widefield Array ($\sim$170-200 MHz). The low-frequency data reveal a previously unreported feature in the average profile (at 185 MHz) after accounting for the effects of temporal broadening arising from multi-path scattering due to the Interstellar Medium (ISM). We advance a new method for flux density calibration of beamformed data from the uGMRT and use it to measure the single pulse flux densities across the uGMRT band. Combined with previously published measurements, these flux densities are best fit with a power-law spectrum with a low-frequency turnover. We also use calibrated flux densities to explore the relationship between pulse-to-pulse variability and the spectral index of individual pulses. Our analysis reveals a large scatter in the single-pulse spectral indices and a general tendency for brighter pulses to show a steepening of the spectral index. We also examine the frequency-dependence of the pulse-fluence distribution and its relation to the Stochastic Growth Theory.

\end{abstract}

%% file: Introduction.tex
\section{Introduction}

Pulsars are rotating neutron stars which are the sites of highly energetic physical processes owing to their extreme environment of strong gravitational and magnetic fields, and high densities. The impeccable periodicity of pulsars was one of the qualities that led to their discovery \citep{1968Natur.217..709H}. Their rotation causes highly collimated radiation beams to sweep across the observer's line of sight for only a brief amount of time, producing a pulsed emission \citep[e.g.,][]{1977puls.book.....M}. Radio emission from pulsars is conjectured to originate within the open field lines in the magnetosphere, with the emission beam centred on the magnetic axis \citep[e.g.,][]{1969ApL.....3..225R}. However, despite a wealth of observational data, the emission mechanism remains not well understood. 

Radio pulsars are known to exhibit a variety of amplitude modulation in single pulses, owing to both intrinsic and extrinsic factors. Even though the integrated pulse profile stays stable over the long term, individual pulses show considerable variation in intensity and shape on a wide range of time scales. The intrinsic variabilities affect pulsar emission at multiple time scales. At the shortest time scales ($\sim\mu$s), pulsars reveal single pulse variabilities such as micro-structures \citep[e.g.,][]{1968Natur.218.1122C}, giant pulses \citep[e.g.,][]{doi:10.1126/science.162.3861.1481, 1970Natur.226..529H, 2003Natur.422..141H}, and giant micro pulses \citep[e.g.,][]{2001ApJ...549L.101J}. At longer time scales of seconds to minutes, the variability is seen in the form of drifting sub-pulses \citep[e.g.,][]{1968Natur.220..231D, 2022MNRAS.509.4573J}, where, as the pulsar rotates, substructures within the pulse are seen to form a regular drifting pattern with longitude. On similar time scales, pulsars also show `moding' where they switch between multiple emission or drifting modes,  average pulse profile, intensity, polarisation, etc. \citep[e.g.,][]{1970Natur.228.1297B, 2013ApJ...775...47C}. On the scale of a few seconds to several minutes, cessation of emission from some pulsars is observed for a certain amount of time, known as nulling \citep[e.g.,][]{1970Natur.228...42B} All these features comprise the intrinsic variability behaviour of single pulses.

At radio frequencies, apart from the intrinsic activity of the pulsar, propagation effects also cause inter-pulse variability \citep[e.g.,][]{1970ApJ...162..707R, 10.1093/mnras/153.3.337}, in the form of extrinsic factors. The interstellar medium (ISM) plays a role in the  modulation of pulsar signals that we observe. Propagation of radio pulsar emission through the ISM causes scattering of the signal and shows up as scintillation on the different timescales of up to tens of minutes \citep[e.g.,][]{1969Sci...166.1401L, 1971MNRAS.155...51S}.

The time variable plasma processes in the pulsar magnetosphere can be most meaningfully investigated using the analysis of single pulses. As mentioned earlier, these individual pulses are affected by the intrinsic and extrinsic factors leading to significant amplitude modulation, which often have a frequency dependence. Studying the single pulses at multiple frequencies allow us to not only sample different regions of the magnetosphere (Radius to Frequency Mapping, \cite{1978ApJ...222.1006C}), but also understand the energy variation with time and frequency, ultimately providing clues to the possible emission mechanisms. Thus, multi-frequency studies of single pulses can be used to investigate the emission properties of the radiation processes in the magnetosphere. A detailed study of these individual pulses can yield constraints on the theoretical models of emission and the distribution of emitting particles, bringing useful insights into pulsar emission physics.

The Stochastic Growth Theory (SGT) has been used to explain the observed pulse energy distribution of several pulsars \citep[e.g.,][]{1995PhPl....2.1466R, 2003ASPC..302..191C, 2003MNRAS.343..512C, 2003MNRAS.343..523C, 2004MNRAS.353..270C}. Theories such as the SGT, help understand the possible emission mechanisms that may give rise to the observed log-normal statistics of single pulse intensities. Some of these aspects can be meaningfully investigated using simultaneous single-pulse observations spanning a large range in frequency. Observational investigations by \cite{2003A&A...407..655K} studied the single pulse behaviour of pulsars PSR B1133+16 and PSR B0329+54 at multiple simultaneous frequencies from 200 MHz to 5 GHz. Their results shed light on the frequency-dependent behaviour of pulse-to-pulse modulation properties and the asymmetric distribution of single pulse spectral indices. Another aspect is the prevalence of log-normal type statistics across a wider population of pulsars. Using the data from the High Time Resolution Universe survey, \cite{2012MNRAS.423.1351B} studied the single pulse energetics and modulation of 315 pulsars at 1352 MHz. They found that over 40\% of the observed sample exhibited a log-normal pulse energy distribution, only a few displayed a Gaussian pulse energy distribution, while the others were not fitted by either distribution. Single pulse analysis of this kind provides new avenues for looking at the instantaneous plasma activity in the pulsar magnetosphere, and such studies have the potential to yield valuable clues to understanding the spectral behaviour of single pulses and the overall pulsar emission mechanism.

Single pulse studies of pulsars necessitate exceptionally high-quality data. The main requirements for such an investigation are - (1) the pulsar should be bright such that single pulses are detectable, (2) low Radio Frequency Interference (RFI) conditions, (3) single pulses should be free of other phenomenology such as nulling or moding. Often, a combination of these conditions is hard to achieve, thus limiting such studies. PSR J1820--0427 is a bright ``normal'' pulsar, with a period of 0.598 seconds and a dispersion measure (DM) of 84.435 pc cm$^{-3}$. The pulsar has a feature-less profile at low frequencies and does not show any kind of modulation (e.g., subpulse drifting, moding, or nulling). Such typical character of the source, makes PSR J1820--0427 an attractive candidate for studying the emission and basic characteristics of pulsars. 

Single pulse studies have been rare in the past due to limited telescope sensitivity. Pulse-to-pulse variability and pulse fluence analyses of pulsars have often evaded attention, mainly because of low single pulse signal-to-noise ratio (S/N). However, such studies can provide a unique instantaneous window into the pulsar magnetosphere and are thus important in gaining insight into the pulsar emission process. In this work, we present a single pulse analysis of the pulsar PSR J1820--0427 using high-quality data obtained from the upgraded Giant Meterwave Radio Telescope (uGMRT) and the Murchison Widefield Array (MWA). In this paper, we investigate the pulse-to-pulse variability and their frequency dependence to study pulsar emission further. The details of observations used in this work are mentioned in section \ref{sec:observations}. The phased array processing, imaging analysis, and single pulse calibration are described in section \ref{sec:dataprocessing}. This section also outlines the novel approach we have developed to measure single pulse flux densities. These calibrated single pulse flux densities are further utilised to study the pulse-to-pulse modulation through investigations of spectral index variation, fluence distribution, etc. We have also studied the average emission characteristic of J1820--0427 and their evolution at simultaneously observed frequencies. The analysis and results from our study are elaborated in section \ref{sec:analysis}. We have discussed the implications of our results in section \ref{sec:discussion}, and section \ref{sec:conclusion} presents a summary of our work.

%% file: Observations.tex
\section{Observations} \label{sec:observations}

Observations of PSR J1820--0427 were carried out using the uGMRT and the MWA at multiple epochs. The uGMRT was used in a dual-band phased-array mode to simultaneously cover the frequency range of 300–500 and 550-750 MHz. The MWA  observations were made  in the 170-200 MHz band. Thus, the pulsar was observed over a frequency range from 170 to 750 MHz, i.e. more than two octaves.  A summary of observational details is presented in Table \ref{tab:obsdetails}, and further details are summarised below.

\begin{table*}
\begin{tabular}{|c|c|c|c|c|c|c|c|c|c|}
\hline
MJD   & Receiver & Frequency (MHz) & $\Delta$f (kHz) & $\Delta$t ($\mu$s) & $\Delta t _{\rm eff}$ (ms) & $N_{Ant}$ & $N_{Chan}$ & $N_{Pulses}$ & Obs. Time (min) \\ \hline
59009 & uGMRT    & 300-500 &    48         &    327.68  &   0.43 - 1.25     & 13         & 4096        & 6020  & 60        \\ 
59009 & uGMRT    & 550-750 &    48         &    327.68  &   0.33 - 0.39     & 13         & 4096        & 6020  & 60       \\ 
59019 & MWA      & 170-200 &    10         &    100     &   1.43            & 128        & 3072        & 4515  & 45        \\ 
59019 & uGMRT    & 300-500 &    48         &    327.68  &    0.43 - 1.25    & 13         & 4096        & 6020  & 60        \\ 
59019 & uGMRT    & 550-750  &   48         &    327.68  &   0.33 - 0.39     & 13         & 4096        & 6020  & 60       \\ \hline
\end{tabular}
\begin{tablenotes}
    \small
    \item $\Delta$f is the frequency resolution of observation
    \item $\Delta$t is the time resolution of observation
    \item $\Delta t _{\rm eff}$ is the effective time resolution
    \item $N_{Ant}$ is the number of antennas (tiles for MWA) used for the observation
    \item $N_{Chan}$ is the number of frequency channels used in the observation
    \item $N_{Pulses}$ is the number of pulses recorded in the observation
\end{tablenotes}
\caption{Table of observations}  
\label{tab:obsdetails}
\end{table*}

\subsection{The uGMRT}

The Giant Metrewave Radio Telescope (GMRT) is a radio interferometric array with 30 antennas, each of 45-meter diameter, spread over an area of 28 km$^2$ in a Y shape \citep{1991CuSc...60...95S}. After a recent upgrade, the system was equipped with wide-band receivers and digital instrumentation to provide a near-seamless coverage in frequency from 120 MHz to 1600 MHz \citep{2017CSci..113..707G, 2017JAI.....641011R}. In the phased array mode, the incoming signals from all antennas are combined coherently to achieve maximum sensitivity for pulsar observations. The observations were made simultaneously at Band 3 (300-500 MHz) and Band 4 (550-750 MHz) of the uGMRT, where only the total intensity was recorded. For these observations, 13 antennas from the central square and the arms were configured at Band 3. For Band 4, 13 antennas from the Y-shaped arms were used, therefore using the largest baselines. The observations were made at two epochs, $\sim$ 10 days apart, simultaneously at both frequency bands in the identical configuration. Along with the phased array observations, the uGMRT also recorded visibilities that can be used to image the pulsar field. At each epoch, the pulsar was observed for roughly one hour, giving close to 6000 pulses per epoch (see Table \ref{tab:obsdetails}).

\begin{figure}
    \centering
    \includegraphics[width=\columnwidth]{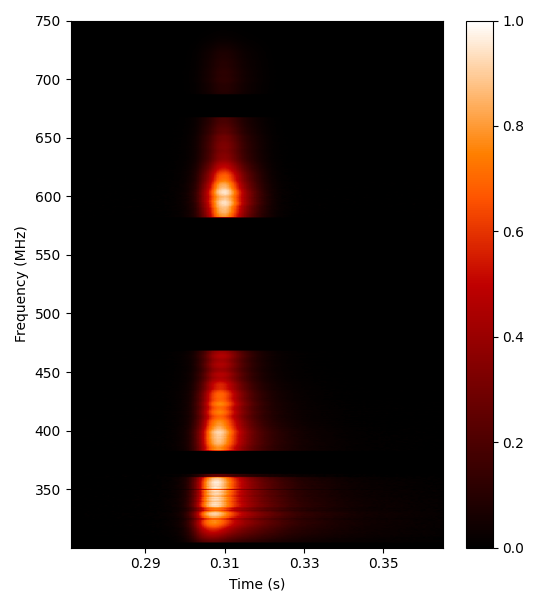}
    \caption{Wide-frequency spectral evolution of PSR J1820--0427 observed using uGMRT. The figure combines Band 3 (300-500 MHz) and Band 4 (550-750 MHz) observations, with frequency increasing from bottom to top. The colour contrast shows the intensity of the pulsar. Apart from the range between 500-550 MHz, the missing chunks correspond to the persistent RFI channels and are removed automatically.}
    \label{figure:freqevol}
\end{figure}

In our uGMRT observations, the PSR J1820--0427 showed very bright radio pulses, along with low RFI conditions, which made this observation suitable for single pulse analysis. The wide bandwidth of uGMRT observations and the impeccable signal-to-noise ratio resulting from its high sensitivity is evident from Fig. \ref{figure:freqevol}. Here, the $y-$axis shows the increasing frequency scale from 300-750 MHz, and the $x-$axis shows a part of the entire pulsar period. All the intensities are normalised to adjust the colour scale in the 0-1 range. Band 3 of uGMRT covers 300-500 MHz and Band 4  550-750 MHz, and an empty region of zero intensity was added in the 500-550 MHz range. The signal strength changes as a function of frequency due to the pulsar's spectrum, and moreover the pulsar profile evolution with frequency can also be noted, where the profile seems to get wider with decreasing frequency.

\subsection{The MWA}

The MWA observations were made at a central frequency of 185 MHz. The MWA comprises 128  tiles, (where a `tile' is a $4m \times 4m$ dipole array) and during our observations, operated in Phase-2 extended array configuration \citep{2013PASA...30....7T, 6b4eee883b0c47faadb947f7fc5ed712}. The data were recorded using the Voltage Capture System (VCS, \cite{2015PASA...32....5T}), which records Nyquist-sampled dual polarisation voltages. In this mode, the channelised voltages are recorded from each tile, which can be processed offline using the software tied-array beamformer \citep{2015PASA...32....6O}, by summing coherently after calibration. We observed the pulsar in full polarisation mode at 185 MHz over a contiguous bandwidth of 30.72 MHz, spread across 3072 $\times$ 10 kHz fine channels. The recorded voltages can also be used to create visibilities, and further obtain the images and source flux at 185 MHz. PSR J1820--0427 was one of the 18 pulsars detected in a single observation under the project G0071, the data from which have been utilised for different projects including developing image-based methods for identifying pulsar candidates \citep{2022arXiv221206982S}, and for investigating the ionospheric refraction offsets in tied-array beamforming \citep{2022PASA...39...20S}. Details on these pulsars, including their profiles and flux densities are reported in \cite{smart1, smart2}. For the data used in this work, the pulsar was observed for 45 minutes, giving close to 4500 pulses per epoch.

%% file: DataProcessing.tex
\section{Data Processing} \label{sec:dataprocessing}

As described earlier, observations were made such that both beamformed and visibility data can be extracted from both, MWA and uGMRT. In the following sections, we describe the procedures adopted for data processing for both phased array and imaging data from these two telescopes. 


\subsection{uGMRT}
\subsubsection{Imaging}
The visibility data  were recorded in the Long Term Accumulation (LTA) format, and were processed using \texttt{SPAM} (Source Peeling and Atmospheric Modeling, \cite{2014ASInC..13..469I}) pipeline and \texttt{CASA} (Common Astronomy Software Applications, \cite{2007ASPC..376..127M}) for calibrating the single pulses obtained from the beamformed data. The \texttt{CASA} package is a data processing package, used to process data from both single-dish and aperture-synthesis telescopes. \texttt{SPAM} is radio astronomy data processing software developed by  \citet{2009A&A...501.1185I} (see also \citealp{2014ascl.soft08006I}), which is a fully automated software based on the Astronomical Image Processing System (AIPS, \cite{1998ASPC..145..204G}). \texttt{SPAM} includes direction-dependent calibration, modelling, and imaging for correcting ionospheric dispersive delays. It uses ParselTongue interface \citep{kettenis2006parseltongue} to access \texttt{AIPS} tasks, files, and tables. \texttt{SPAM} consists of two main processing steps: (1) a pre-processing part that converts raw data from individual observing sessions (LTA format) into pre-calibrated visibility data sets, and (2) a main pipeline which converts pre-calibrated visibility data into stokes I continuum image.

\begin{figure*}
    \centering
    \includegraphics[width=\textwidth]{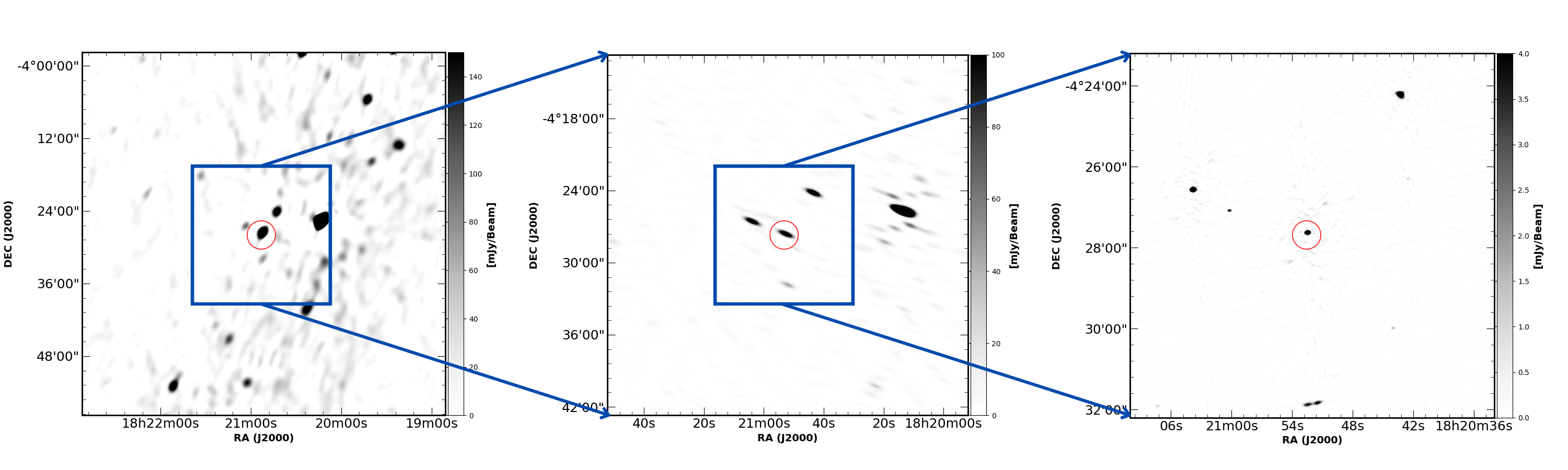}
    \caption{\textit{Left}: MWA Stokes I image of the $1\times1$ deg$^2$ field (at 185 MHz) with the position of the target pulsar in the centre, with the rms sensitivity is $\sim500$ $\mu$Jy/beam. \textit{Middle}: Combined uGMRT Band 3 (300-500 MHz) image of $0.5\times0.5$ deg$^2$ field, where the rms sensitivity is $\sim300$ $\mu$Jy/beam. \textit{Right}: Combined uGMRT in Band 4 (550-750 MHz) image of $0.15\times0.15$ deg$^2$ field, with a much higher sensitivity of $\sim100$ $\mu$Jy/beam. In all the images pulsar is at the phase centre, as indicated by the red circle.}
    \label{figure:image}
\end{figure*}

We used the \texttt{SPAM} pipeline to obtain the Stokes I continuum image. After converting LTA files in the \texttt{.UVFITS} \footnote{a FITS file with telescope metadata and visibility information} format, \texttt{SPAM} derives the calibration and flagging information from the primary calibrators. These solutions are then applied to the target. The data were automatically split into smaller subbands by \texttt{SPAM}, such that the primary beam for each subband does not vary substantially withing the frequency range. \texttt{SPAM} automatically splits the Band 3 data into six equal subbands, whereas Band 4 data was split into four subbands of equal bandwidth. These subbands were further used to study the frequency evolution of the pulsar emission.

\input{table1.tex}

Each sub-band was processed individually, keeping the reference frequency same for all the subbands created from one band. For Band 3, the reference frequency was 400 MHz and for Band 4 it was 650 MHz. The targets in each subband are then calibrated, where the different target fields are separated. In the main \texttt{SPAM} pipeline run, the representative resolution of the model image was set as $10''$ for Band 3 subbands and $6''$ for Band 4 subbands. Out of the total six subbands in epoch 1 Band 3 observation, only five were usable. However, for epoch 2, Band 3, we had to discard data from two subbands due to poor data quality. The poor data quality in these cases is due to RFI being limited to certain frequencies. Thus, subbands where most of the data were flagged by \texttt{SPAM} were discarded, whilst subbands with no persistent or frequency-limited RFI were used in further analysis. Each \texttt{SPAM} pipeline run on a sub-band yielded a final image and a calibrated visibility data set (\texttt{.SP2B.CAL.RR.UVFITS}). Imaging was performed on the final calibrated data using the \texttt{CASA} task \texttt{tclean}. To get the flux density of PSR J1820--0427, we used the polygon drawing option in \texttt{casaviewer} to roughly make a polygon around the source of interest and then fit the selected region using a Gaussian to get the integrated flux density value.

As seen from Fig. \ref{figure:image}, our analysis attains S/N of $\sim$1000 for simultaneous imaging detection of this pulsar. The observation details and imaging results for all the observed frequencies are summarised in Table \ref{tab:image}. An image of the pulsar field at Band 3 and 4, combining all the subbands, is shown in Fig. \ref{figure:image}. The Band 3 image (middle) shows a 0.5-degree cropped region around the pulsar, and Band 4 image (right) shows a 0.15-degree region around the pulsar. In the images, pulsar is at the phase centre, denoted by a red circle. The colour bar shows the associated flux density scale in mJy/beam. The Band 3 image in Fig. \ref{figure:image} is visibly elongated, as compared to the Band 4 image. This is because the telescope configuration used for Band 3 observations comprised primarily antennas located within the central square plus a few additional antennas, thus covering predominantly shorter baselines ($\lesssim$1\,km). However, Band 4 observations were made using antennas located along the three arms that make the Y-shaped array, and hence covering mostly longer baselines ($\sim$1-25\,km).

\subsubsection{Phased Array Beam}

The uGMRT phased array data were recorded with a time resolution of 327.68 $\mu$s and a frequency resolution of 48.83 kHz. The 200 MHz observing band was split into 4096 frequency channels. The resulting data were written in the filterbank format and subsequently converted to single pulse archives using the {\tt DSPSR} package \citep{2011PASA...28....1V}.

Each single pulse archive file was split into multiple subbands that were identical to those used in the \texttt{SPAM} pipeline, using the {\tt psrsplit} subroutine of {\tt PSRCHIVE} \citep{2004PASA...21..302H}. For Band 3 observations, all single pulse archives were split into six subbands, with centre frequency matching the {\tt SPAM} subband files. Similarly, single pulse archives from Band 4 observations were divided into four subbands. Single-pulse archives of each frequency subband were then frequency averaged and combined using the \texttt{pam} and \texttt{psradd} routines from {\tt PSRCHIVE}. Data corresponding to the unusable \texttt{SPAM} subbands were discarded. The phased array observation made using uGMRT showed a total S/N of close to 10,000, thereby yielding a mean single pulse S/N of $\sim$100. The data were also manually searched for stray RFI occurrences using the interactive RFI zapping subroutine {\tt pazi} of {\tt PSRCHIVE}. The RFI-excised file was then converted into an ASCII format which contained the single pulse time series. This time series was subsequently calibrated in units of Jansky (Jy), the details of which are described in section \ref{sec:fluxcalib}.


\subsection{MWA} 
\subsubsection{Imaging}
The xGPU software correlators \citep{2013IJHPC..27..178C} were used to create ``visibilities'' at 1s time resolution from the raw voltages recorded with the VCS. These visibilities were then converted into {\tt CASA} measurement set format \citep{2007ASPC..376..127M} using the {\tt COTTER} software \citep{2015PASA...32....8O}. The application of calibration and the removal of RFI affected channels using {\tt AOFlagger} \citep{2012A&A...539A..95O} were also done during this conversion. The calibration solutions were obtained from the Murchison Widefield Array All-Sky Virtual Observatory (MWA-ASVO, \citet{2020PASA...37...21S}). {\tt WSCLEAN} \citep{2014MNRAS.444..606O} was used to form images in the instrumental polarisation which were then converted to Stokes IQUV images using the MWA full embedded element beam model \citep{2017PASA...34...62S}. The images generated were 8192 $\times$ 8192 pixels, with a pixel size of 0.049 degree/pixel, producing $40^\circ \times 40^\circ$ images. The mean standard deviation of the Stokes I image used for processing is $\sim 5$ mJy/beam around the centre of the image (see \cite{2022arXiv221206982S} for further details). For PSR J1820--0427, the estimated flux density at 185 MHz was 790 $\pm$ 63 mJy. Fig. \ref{figure:image} (left) shows the $1\times1$ deg$^2$ region around the pulsar (red circle) at 185 MHz.

\subsubsection{Phased Array Beam}

The MWA phased array data were generated with a time resolution of  100$\mu$s and a  frequency resolution of 10\,kHz. The 30.72-MHz observing band was centred at 185 MHz. For the MWA tied-array beamformer, signals from each antenna were recorded by the MWA VCS. The voltages from all antenna elements are then calibrated and combined coherently \citep{2019PASA...36...30O}, in offline processing on the Pawsey supercomputer. The resulting data were written in the {\tt PSRFITS} format \citep{2004PASA...21..302H} and were subsequently converted to single pulse archives using the {\tt DSPSR} \citep{2011PASA...28....1V} and {\tt PSRCHIVE} \citep{2004PASA...21..302H} package. The data were exceptionally clean and required no RFI excision, however, the mean single pulse S/N $\sim$10, and therefore the data quality is not good enough to perform a detailed single pulse analysis. The single pulse archives were thus time scrunched, thus giving an average profile, which was further converted into ASCII format for further analysis.



\subsection{Flux Density Calibration}\label{sec:fluxcalib}

Flux density calibration for pulsar observations generally relies on the use of noise diodes, which are typically available for most single-dish telescopes such as Parkes (Murriyang). Even though the GMRT antennas are equipped with switchable noise diodes, their synchronisation across the array is currently not a functional mode for  the uGMRT. We have therefore developed a new method that takes advantage of the fact that concurrent imaging is possible with uGMRT while recording beamformed (phased array) data. Basically, visibility data are imaged first to obtain calibrated images (in units of Jy) which are then used to bootstrap to achieve a first order calibration of the beamformed data in flux density units.

The underlying assumption is that the sensitivity of the telescope is comparable in both modes of operation: i.e. phased array and imaging interferometer. This would necessarily imply that the RMS noise of the receiver is identical for the beamformed pulsar detection using the phased array beam and an imaging detection using the cross-correlated visibilities. Therefore, using the off-source RMS from each image (see Table \ref{tab:image}) and applying a suitable scaling factor, we have calculated the equivalent RMS of the off-pulse region (in the phased array mode) of the average pulsar profile as  
\begin{equation} \label{eqn:boostf}
     \sigma_{\text{\rm PA}} = \sqrt{n_{\rm binoff}} \times \sigma_{\text{\rm Img}}
\end{equation}
where $\sigma_{\rm PA}$ is the RMS value in the phased array mode, $\sigma_{\rm Img}$ is the image RMS, and $n_{\rm binoff}$ is the number of bins in the off-pulse duration. Using the average pulsar profile for each frequency, the on-pulse region was defined as where the pulse intensity drops to less than 5\% of the peak intensity. The quantity $n_{\rm binoff}$ was then calculated by subtracting the number of bins in the on-pulse region from the total number of phase bins in the pulse profile. This scaling factor accounts for a fraction of the pulsar period during which no emission was seen from the pulsar, i.e., when the pulsar was ``off'' in its duty cycle. Further, according to our assumptions, this $\sigma_{\text{PA}}$ should be the same as the RMS noise of the off-pulse region in the averaged profile (arbitrary units). Therefore, we calculated the conversion factor $\sigma_{\text{PA}}$/$\sigma_{\text{offpulse}}$ and applied it to our analysis to bring the beamformed time series data in physical units (Jy). As a useful cross-check, we calculated the mean flux density from the average profile of the scaled beamformed data. This number, in principle, should be equal to the flux density obtained from the imaging analysis, assuming various factors that impact or alter the sensitivity achievable in practice are accounted for (e.g. some degree of de-phasing of the array that may be caused over time). Barring a few outlier cases, the calculated flux density values were found to be in $\sim 20-30\%$  agreement with the measured flux density values. We have therefore used these calibrated time series to perform all the analyses henceforth.


%% file: table1.tex
\begin{table}
\centering
\begin{tabular}{|c|c|c|c|c|}
\hline
\multicolumn{1}{|c|}{Frequency} & \multicolumn{1}{c|}{Bandwidth} & \multicolumn{1}{c|}{Flux $\pm$ Error} & \multicolumn{1}{c|}{Off-source RMS} & \multicolumn{1}{c|}{Restoring} \\
\multicolumn{1}{|c|}{(MHz)} & \multicolumn{1}{c|}{(MHz)} & \multicolumn{1}{c|}{($m$Jy)} & \multicolumn{1}{c|}{($\mu$Jy)/beam} & \multicolumn{1}{c|}{Beamsize} \\ \hline
185 & 30.72 & 790 $\pm$ 63 & 500 & 83$''$ $\times$ 55$''$ \\
317 & 33 & 256.58 $\pm$ 2.06 & 360 & 73$''$ $\times$ 25$''$ \\
350 & 33 & 206.15 $\pm$ 0.85 & 340 & 66$''$ $\times$ 23$''$ \\
383$^a$ & 33 & 132.80 $\pm$ 1.30 & 380 & 60$''$ $\times$ 21$''$ \\
417 & 33 & 145.84 $\pm$ 0.14 & 110 & 55$''$ $\times$ 19$''$ \\
450 & 33 & 125.87 $\pm$ 0.45 & 120 & 51$''$ $\times$ 18$''$ \\
483 & 33 & - & - \\
575 & 50 & 76.43 $\pm$ 0.53 & 110 & 5$''$ $\times$ 3$''$ \\
625 & 50 & 62.60 $\pm$ 0.38 & 80 & 4$''$ $\times$ 3$''$ \\
675 & 50 & 50.30 $\pm$ 0.37 & 80 & 4$''$ $\times$ 3$''$ \\
725 & 50 & 41.96 $\pm$ 0.38 & 80 & 4$''$ $\times$ 2$''$ \\ \hline
\end{tabular}
\begin{tablenotes}
    \small
    \item $^a$ Only for epoch 1
\end{tablenotes}
\caption{Multi-frequency flux density measurements of \src obtained from imaging the pulsar field using MWA (185 MHz) and uGMRT(317-725 MHz). The flux densities and their respective errors are added in quadrature for multiple epoch observations. Data at 483 MHz was unusable and was discarded for this work. At Epoch 2, data at 383 MHz was unusable, thus only the values from Epoch 1 were used.}
\label{tab:image}
\end{table}

%% file: Analysis.tex
\section{Analysis and Results} \label{sec:analysis}

\subsection{Profile Evolution}

\begin{figure}
    \centering
    \includegraphics[width=\columnwidth]{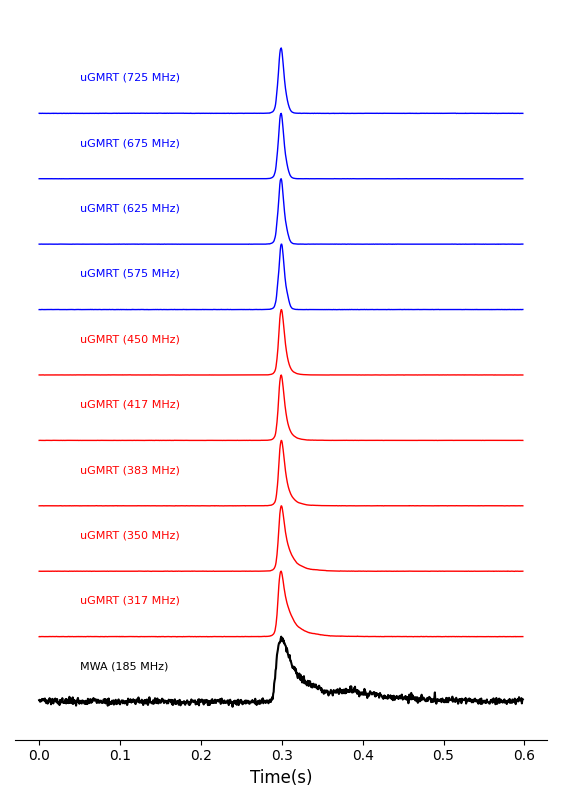}
    \caption{Normalised average profiles of PSR J1820--0427 at all the observed frequencies from 185 MHz to 750 MHz. The 185 MHz observation was made using MWA, while the rest were made using uGMRT. The MWA profile has 512 time bins, whereas the uGMRT profiles are made with 1024 time bins. In the figure, $x-$axis is the time scale, with the average profile peaks aligned at the centre. The average profiles made using the uGMRT data show the impeccable data quality with off-pulse intensity close to zero. The scattering tail can also be seen at the lower frequencies, and an extra component rising at 185 MHz towards the trailing edge.}
    \label{figure:avg_wtrfall}
\end{figure}

Fig. \ref{figure:freqevol} shows the time-averaged frequency structure of the pulse profile with phase, which also illustrates the frequency-dependent behaviour of the profile width and the pulse intensity. The profile width increases with decreasing frequency. Similar information is shown in Fig. \ref{figure:avg_wtrfall}, where average profiles of PSR J1820--0427 at multiple observing frequencies are shown. The $x-$axis shows time equivalent to one pulsar period (0.598s), where the peak of every averaged profile for each frequency is centred at the midpoint. The 200\,MHz bandwidth of uGMRT data were divided into several subbands following the exercise in section \ref{sec:dataprocessing}. The Band 3 uGMRT data were divided into six equal subbands, and data at Band 4 were divided into four equal subbands. As shown in Fig. \ref{figure:avg_wtrfall}, the average pulsar profile has a single sharp peak, with a scattering tail most pronounced at the lower frequencies. The excellent quality of the uGMRT data can also be seen clearly in each of the averaged profiles at frequencies from 317 to 725 MHz, where the off-pulse noise is negligible.

The pulsar profile at the MWA frequency band (170-200 MHz) shows a secondary feature near the trailing end of the profile, for which no counterparts are visible in the uGMRT bands. Past observations at higher frequencies \citep{1998A&AS..127..153K, 2019ApJ...874...64Z}  have reported a multi-component profile at frequencies above $\sim1.5$ GHz, however, no such extra components are known at lower frequencies.


\subsection{Pulse-to-pulse Variability} \label{sp_variability}

\begin{figure*}
    \centering
    \includegraphics[width=\textwidth, trim={0.2cm 0cm 0.1cm 0cm}]{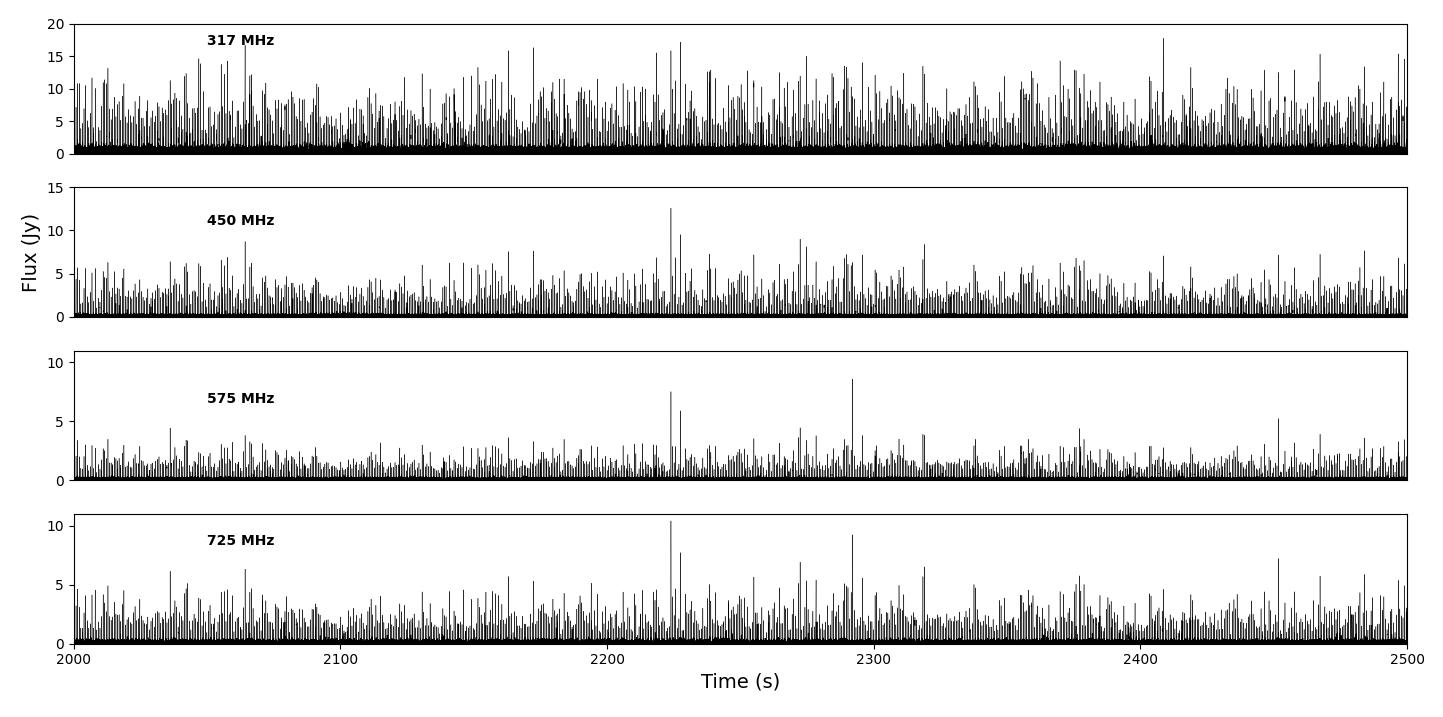}
    \caption{Calibrated single pulse train for PSR J1820$-$0427 at multiple observation frequencies for uGMRT Epoch 1. The high variability of single pulses can be seen, which decreases with higher frequency. A single high amplitude pulse also emerges at higher frequencies (around 2225s), showing a peculiar spectral index.}
    \label{fig:sp_variability}
\end{figure*}

Fig. \ref{fig:sp_variability} shows the pulse train (a sequence of 400 pulses) at  different  frequencies across the uGMRT bands, where pulse-to-pulse variations can be  readily observed.  Amplitude fluctuations in pulsars can arise due to both intrinsic and extrinsic effects. While there seem to be multiple instances of large-amplitude pulses, they do not fit the standard criteria of giant pulses, which requires flux densities of  tens or hundreds of times larger than the average pulsar flux. Moreover, in contrast with previous studies on nulling statistics \citep{1992ApJ...394..574B}, which suggested a small nulling fraction of the order of $<$1\% for this pulsar, our analysis does not reveal any instances of true nulls, despite having over 12000 pulses observed at multiple epochs and frequencies, with each pulse having a S/N of $\sim$100.

The fluctuations in pulse-to-pulse intensities can be quantified in terms of  the modulation index, to further investigate the single pulse variability. Modulation Index ($m$) is defined as - $\sqrt{\langle \Delta I^2 \rangle}/ \langle I \rangle$, where $I$ is the pulse intensity, and the angle brackets represent averaging over a large ensemble of adjacent pulses and can be in computed as a function of the pulse phase, $\phi$. We found the modulation index for this pulsar to be close to 0.5, with variations along the on-pulse phase. This number is slightly higher than the minimum on-pulse, phase-resolved modulation index of $\sim0.34$ calculated in \cite{2012MNRAS.423.1351B}. In the observed frequency range of 300 to 750 MHz, the modulation index shows a maximum near the peak pulse phase and falls off at the tails.

Using the calibrated single pulse data, we further studied the spectral behaviour of individual pulses. We estimated the spectral index for each pulse using the flux density estimates at every two consecutive frequencies, following the relation -
\begin{equation}
    \alpha(n) = \frac{\text{log}[S_{\nu_1}(n) / S_{\nu_2}(n)]}{\text{log}(\nu_1/\nu_2)}
\end{equation}
where n is the pulse number, $S_{\nu}$ is the flux density at frequency $\nu$, and $\alpha$ is the spectral index. Thus, an $\alpha(n)$ value was calculated for every pulse using the flux density at consecutive observed frequencies. The spectral index values obtained in this study were seen to vary drastically from pulse to pulse and over the consecutive frequency pairs. This will be discussed in detail in section \ref{sec:sec5.2}.


\subsection{Pulse Fluence Distribution} \label{pulse_en}

The investigation of pulse energy distribution provides insights into the radio pulsar emission mechanism and the physical state of the pulsar magnetosphere \citep[e.g.,][]{2003ASPC..302..191C,2012MNRAS.423.1351B}. Many of the observed phenomena, including pulse-to-pulse variability (in pulse amplitude, structure, etc) point to the fact that physical conditions in the magnetosphere tend to change on very short time scales, i.e., pulsar emission process is inherently dynamic in nature. Observations such the distribution of single pulse energies can provide a window to the instantaneous state of pulsar plasma and the nature of emission mechanisms that give rise to the observed distribution \citep[e.g.,][]{2012MNRAS.423.1351B}.

For the following exercise, we only used the calibrated uGMRT single pulse data. Single pulses at 185 MHz using MWA had low S/N and could not be used to study the pulse energy distribution, using single pulse data. We visually determined the on-pulse region in each observation using the average profile. The on-pulse region was selected as the width at which the flux density drops to less than roughly 5\% of the maximum flux density in the average profile. The on-pulse energy ($E$) was calculated for every single pulse by integrating the calibrated flux in the on-pulse region. The estimated pulse energy has units of mJy ms, i.e., pulse fluence. A histogram of pulse fluence was then constructed to study the distribution.

\begin{figure*}
    \centering
    \includegraphics[width=\textwidth]{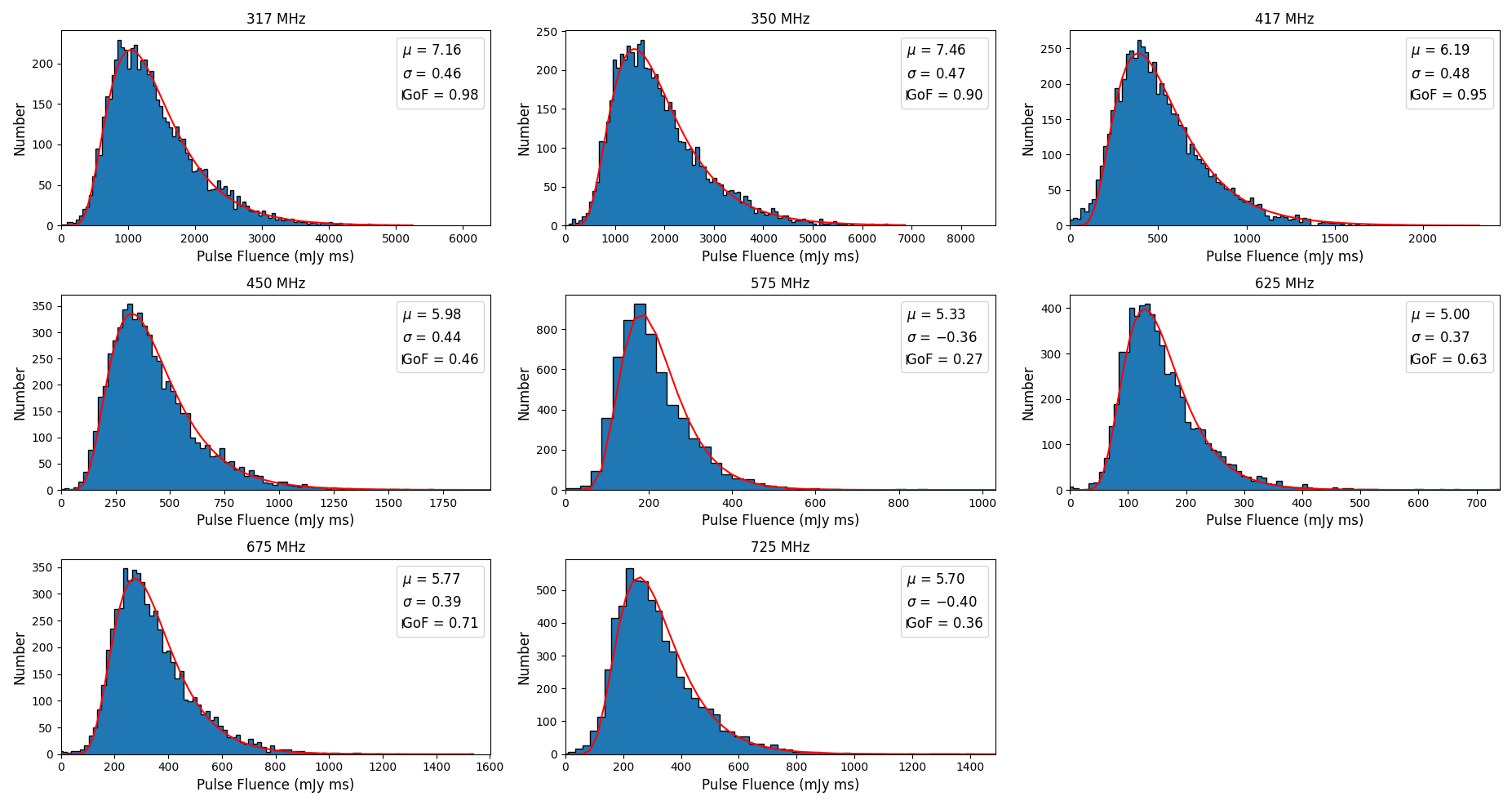}
    \caption{Pulse fluence histograms for uGMRT observations from Epoch 2. Each panel represents the pulse fluence distribution at a frequency mentioned at the top of each panel. In support of theories like the Stochastic Growth Theory (SGT), we found a log-normal distribution of pulse energies consistent at all frequencies. The histogram is fitted with a log-normal distribution shown as the red curve, with the parameters mentioned in the top right corner of each subplot.}
    \label{figure:pulsen_fit}
\end{figure*}

In Fig. \ref{figure:pulsen_fit}, each panel shows the pulse fluence distribution of PSR J1820--0427 at different subband frequencies. It is evident that the pulse energies follow a log-normal distribution at all the observed frequencies. The log-normal distribution is the probability distribution of a random variable whose logarithm is normally distributed. We fitted a log-normal model to the observational data using a least-squares fitting method, defined using the parameters $\mu$ and $\sigma$:
\begin{equation}
    \text{N(E)} = \frac{A}{E} \text{exp}{\left[-\frac{\left( \text{ln} (E) - \mu \right)^2}{2 \sigma^2}\right]}
\end{equation}
where $E$ is the pulse fluence, $A$ is the scaling factor, and $N(E)$ is the log normal distribution of fluence with parameters $\mu$ (mean) and $\sigma$ (standard deviation). We found the log-normal pulse fluence distribution  best describes our data, and over the observed wide-frequency range and both epochs. A goodness-of-fit (GoF) was also calculated for each case to determine the quality of the fit. The parameter $\mu$ and $\sigma$, along with the GoF, are given in Fig. \ref{figure:pulsen_fit}.

Our study extends the findings of \cite{2012MNRAS.423.1351B} at 1.4 GHz, to frequencies below $\sim$1\,GHz, spanning a wide-frequency range of 300-750 MHz, using a much larger sample size of over 12,000 pulses. Our analysis revealed that the emission process giving rise to the log-normal behaviour of pulse energies remains the same over a wide frequency range.


\subsection{Flux density spectrum} \label{sec:spect_ind}

The calibrated pulsar flux densities were calculated over a wide range of frequencies (170-750 MHz) by imaging the pulsar field (see Fig. \ref{figure:image}), as listed in Table \ref{tab:obsdetails}. It is possible that the flux density values at lower frequencies can be less reliable due to inherently larger source confusion at these frequencies and comparatively larger fields-of-view. As visible from the images in Fig. \ref{figure:image}, the pulsar is in the vicinity of multiple sources, which may slightly bias the estimated flux densities at lower frequencies. Yet, even  at 185 MHz, the pulsar is at least $6\sigma$ away from the closest source. Hence, the source is well separated from the close-by sources, thus providing a reliable flux density estimate.

\begin{figure}
    \centering
    \includegraphics[width=\columnwidth]{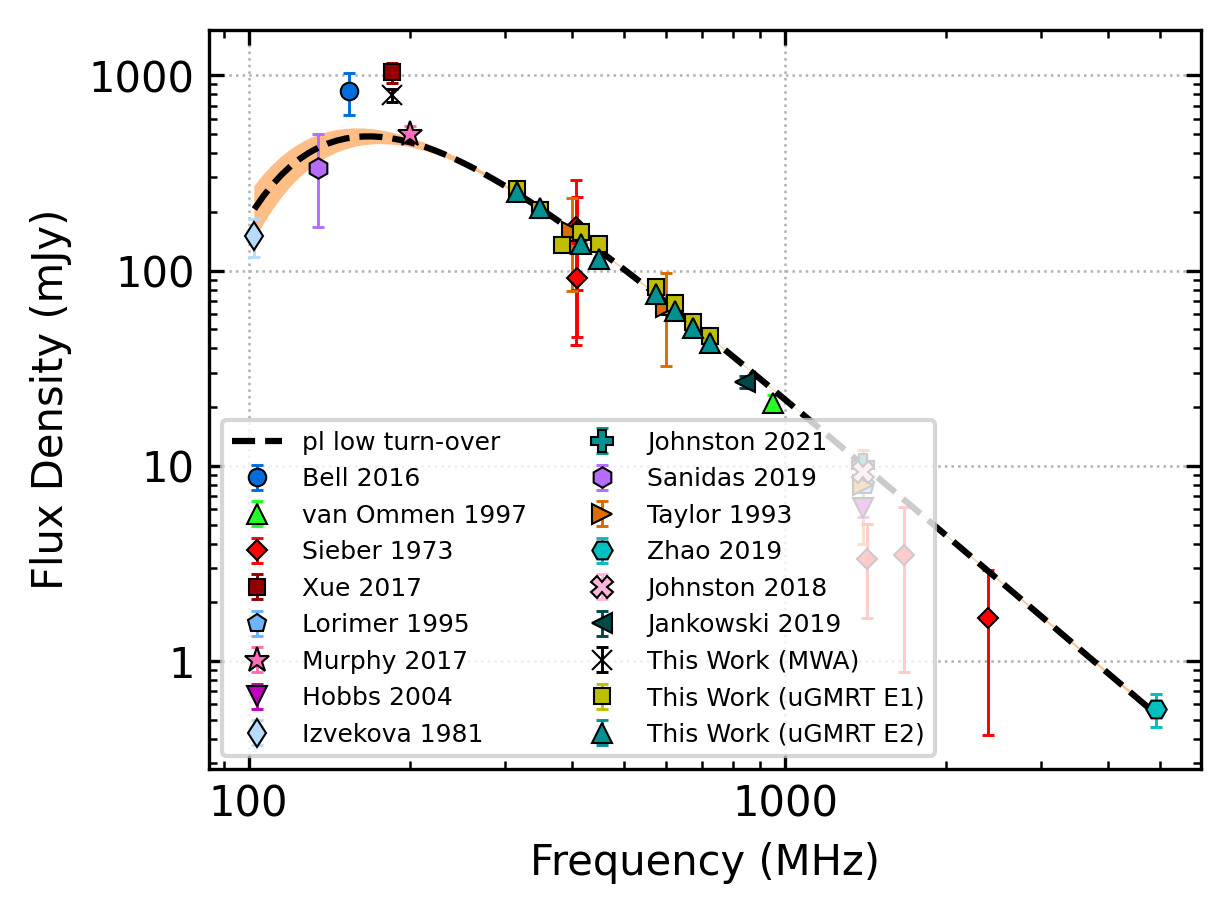}
    \caption{Spectral Index fit for PSR J1820$-$0427 using flux density values from this work along with the flux density values from the published literature. The black dashed line is the best-fitting model to the data, and the orange shaded envelope is the 1$\sigma$ uncertainty of the best-fitting model. The turn-over frequency was found to be $\sim$170 MHz and the spectral index is -2.34$\pm$0.01 In contrast with the previous studies, the power law with a low-frequency turn-over (dashed line) fits the data best for this pulsar.}
    \label{figure:spind}
\end{figure}

We used the \texttt{pulsar\_spectra}\footnote{\url{https://github.com/NickSwainston/pulsar_spectra}} repository \citep{2022arXiv220913324S}, which is a fully-featured spectral fitting PYTHON software package. It also contains a catalogue of flux density measurements from several publications, which were used to collate all the flux density measurements of this source from this work and the existing literature. The resultant spectral fit of PSR J1820--0427 using all known flux densities is shown in Fig. \ref{figure:spind}. The spectral fitting procedure is based on \cite{2018MNRAS.473.4436J}. To account for the uncertainties of the outlier points, a Huber loss function \citep{10.1214/aoms/1177703732} is used instead of the least-squares quadratic loss function. The function deviates to a linear loss once a certain distance is reached from the model. The data are then fitted using different models like a power law, broken power law, log-parabolic, etc. The best fit model is then determined using the Akaike information criterion (AIC, \cite{1100705}), which is a measure of information retained by the model without over-fitting.

In the published literature, PSR J1820--0427 is known to exhibit a broken power law \citep{2017PASA...34...20M,2022PASA...39...42L}. Fig. \ref{figure:spind} combines 17 flux density measurements (multiple epochs) from this study with all other published values from the literature. Our analysis suggests a power-law function with a low-frequency turn-over to best describe spectral behaviour of the pulsar, as shown by the black dashed line in Fig. \ref{figure:spind}. Here the one-sigma uncertainty on the model is given by the orange-shaded envelope. As per  this model, the turn-over frequency is 169$\pm$5 MHz and the spectral index is -2.34$\pm$0.01. Given the spectral turn-over of pulsar, it is possible to reconcile that the single pulses at 185 MHz were not bright enough for a detailed single pulse analysis.


\subsection{Scatter Broadening} \label{scattering}

The pulse profiles in our observations show substantial scatter broadening of the pulsar, which may be expected given the moderate DM and proximity to the galactic longitude and latitude ($l \sim 25^\circ$, $b \sim 5^\circ$). For PSR J1820--0427, we are able to characterise the temporal broadening of the pulse caused by multi-path scattering, and its frequency dependence, given the significant evolution across the observing band from $\sim$170 MHz to $\sim$750 MHz. As seen in Fig. \ref{figure:avg_wtrfall}, scattering is readily visible at the lower frequencies of the uGMRT and most pronounced at the low-frequency band of the MWA. As seen from Fig. \ref{figure:avg_wtrfall}, the pulse profile is minimally scattered at the higher frequencies, but the degree of scattering steadily increases at lower frequencies. Though the pulse profile looks relatively simple at the uGMRT bands, it clearly shows an additional feature in the pulsar profile at the MWA band. This is the first reported detection of this additional feature for J1820--0427.

The scattering of pulsar emission depends on fluctuations in the Inter-Stellar Medium (ISM) electron density and results in the broadening of pulse shape at lower frequencies. This pulse broadening is quantified by the timescale, $\tau_d$, which is characteristic of a pulse broadening function (PBF) fitted to a measured pulse shape. The broadening time has a  frequency dependence, given by  $ \tau_d \propto \nu^{-\beta}$, where the scaling index $\beta$ depends on the details of scattering geometry and the nature of the turbulent medium. The PBF is the impulse response of the ISM to a signal approximated as delta function. The exact form of the PBF associated with the ISM is generally unknown. It depends on the distribution and the turbulence power spectrum of the scattering material along the line of sight to the pulsar. Its frequency dependence is measured by the frequency scaling index, $\beta$, which is related to the underlying physics of turbulence and its power spectrum.

For the estimation of pulse broadening time, $\tau_d$, we use the method developed in \cite{2004ApJ...605..759B} and \cite{2019ApJ...874..179K}. Consideration of this method can be justified given the emergence of a secondary hump-like feature at low frequencies, as revealed in the MWA data. The method makes use of a CLEAN-based algorithm to deconvolve the interstellar pulse broadening from the measured pulse shapes. The CLEAN-based approach offers the benefit that no assumption needs to be made about the intrinsic pulse shape. However, it does involve (as with most other methods) trailing a range of PBFs and uses a set of parameters (i.e. figures of merit) to determine the best fit PBF. As demonstrated in
\cite{2004ApJ...605..759B} and \cite{2019ApJ...874..179K}, in principle, it can also be used to ascertain the shape of PBF that gives a better fit.

\begin{figure}
    \centering
    \includegraphics[width=1.1\columnwidth]{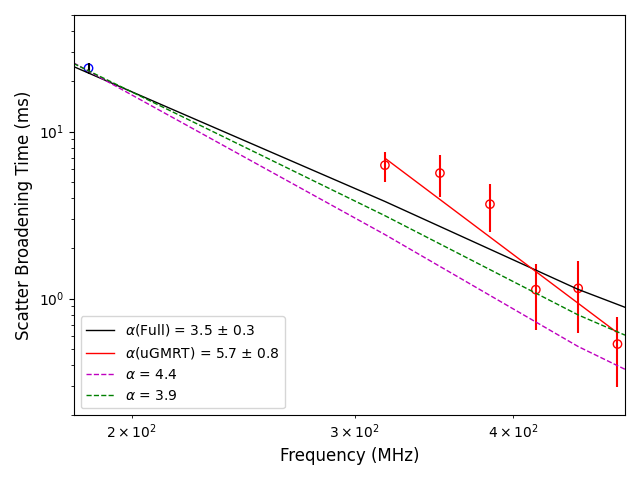}
    \caption{Measured scattering delay vs the observing frequency for PSR J1820--0427; using a thick screen model for scattering, as described in the text. The black line in the plot is a power-law fit of the form $\tau_d \propto \nu^{- \beta}$, to the complete observed data. The red line shows a fit to only uGMRT Band 3 data. The green dashed line shows the frequency scaling for Kolmogorov turbulence, and the blue dashed line is for the empirical index favoured by a large body of measurements as shown by \citet{2004ApJ...605..759B}.}
    \label{figure:sctrbroad}
\end{figure}

For PSR J1820--0427, an exponential PBF with a rounded shape (thick scattering screen) results in better-reconstructed profiles, than the simple exponential that corresponds to an ideal thin screen model. \cite{1973MNRAS.163..345W} showed that for a thick scattering screen, the PBF during the rise time up until shortly after the peak of the emission, is given by
\begin{equation}
    PBF(t) = \left( \frac{\pi \tau_d}{4t^3} \right)^{1/2} \exp \left[ - \frac{\pi^2 \tau_d}{16t} \right]
\end{equation}

Fig. \ref{figure:sctrbroad} shows the measured scatter broadening times  with observed frequencies with different power law fits. The fitting exercise was performed using the MWA data at 185 MHz and uGMRT Band 3 data from the first epoch of observations. The black line shows a power-law fit using the data at 185-500 MHz, with a frequency scaling index, $\beta \sim 3.5 \pm 0.3 $. This is shallower than the theoretically expected value of $\beta$ = 4.4 for a Kolmogorov-type turbulence \citep{1990ARA&A..28..561R}. The overall scaling index is much closer to, albeit lower, than the inferred frequency scaling index from a global fit using a large set of measurements ($\tau_d \propto \nu^{-3.9}$), as shown in \cite{2004ApJ...605..759B}. However, the scaling index, as constrained by uGMRT measurements alone (red line), is much steeper, with an estimated value of $\beta = 5.7 \pm 0.8 $ for $\tau _d$ measurements across the 300-500 MHz range.

As evident from Fig. \ref{figure:sctrbroad}, the measured broadening time at the MWA frequency (185 MHz) is significantly below the expectation based on extrapolation from scaling suggested by uGMRT data. A plausible explanation is perhaps a truncated scattering screen \citep{2001ApJ...549..997C}, where truncation occurs at $\lesssim$ 300\,MHz. Such a scenario can give rise to truncated images and hence truncated PBFs, as the scattering angle is large enough for the rays from the screen's edges to reach the observer. An observational manifestation of this is a shallower scaling index below the break frequency, i.e. an anomalous frequency scaling as discussed by \cite{2001ApJ...549..997C}, however it is hard to be conclusive given just a single low-frequency measurement, especially considering that our measured profiles are better fit by PBFs of a thick screen rather than that of a thin screen.

\begin{figure}
    \centering
    \includegraphics[width=1.1\columnwidth]{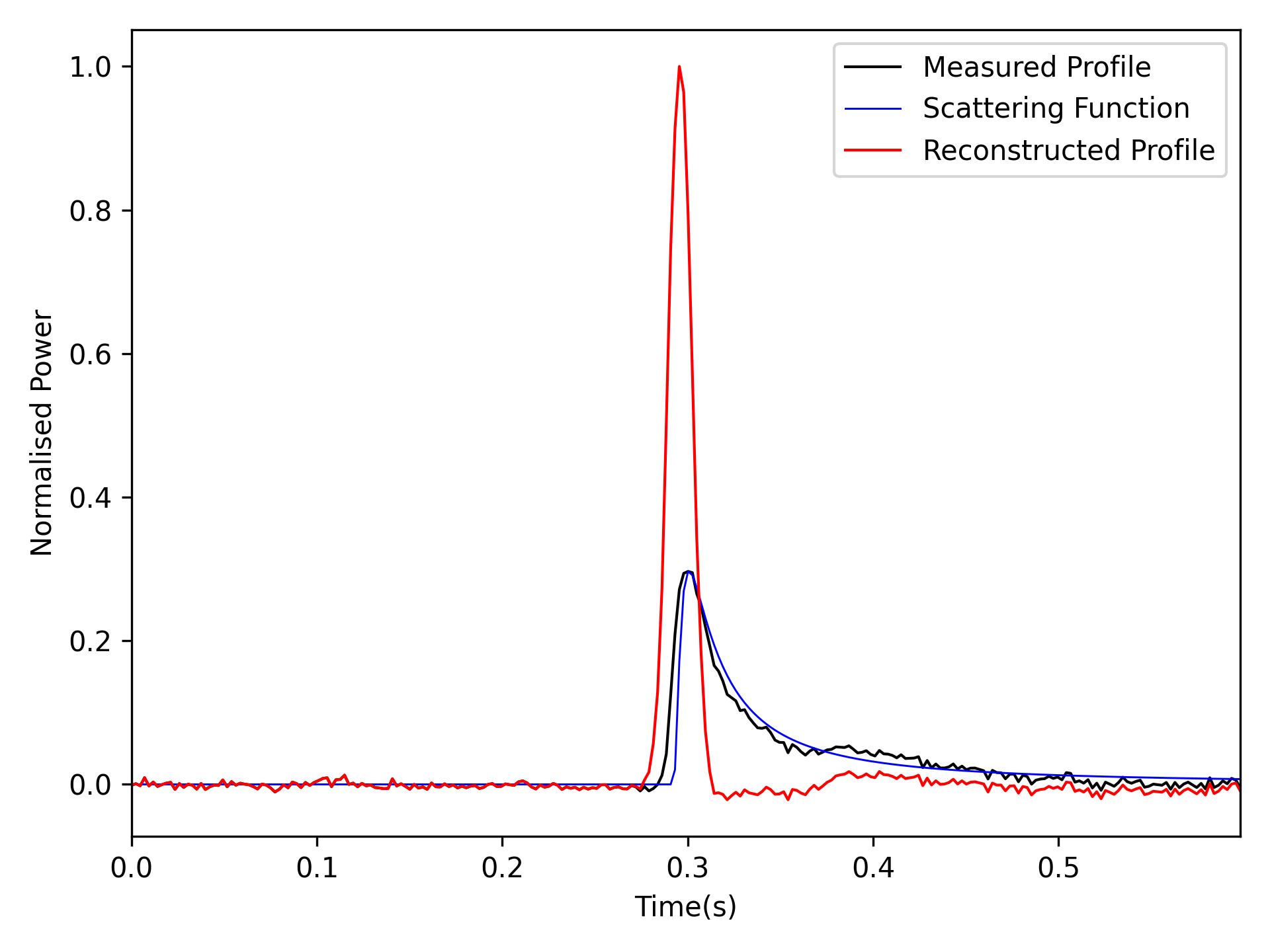}
    \caption{The measured pulse at 185 MHz using MWA is shown in black. The curve in blue shows the scattering function, which was chosen to be an exponential pulse broadening function with a rounded shape (thick scattering screen). The red curve is the reconstructed pulse profile, and clearly shows a secondary feature emerging at the tail end, near time 0.4.}
    \label{fig:Figure8}
\end{figure}

The reconstructed profile at the low-frequency band (185 MHz, MWA), as shown in Fig. \ref{fig:Figure8} in red, reveals a secondary feature akin to post-cursor emission. The presence of the secondary feature can be easily reconciled given the measured pulse profile at 185 MHz (see Fig. \ref{figure:avg_wtrfall}) and is clearly not an artefact of analysis. However, its absence or significantly reduced prominence at the higher frequency bands of uGMRT means either it has a much steeper spectrum than the main feature of the profile, or a new feature that is visible at lower frequencies due to the specific beam/emission geometry of the pulsar. The latter may need a broader frequency coverage at low frequencies for detailed investigation (e.g. observations spanning a more extensive frequency range with the MWA).


\subsection{Faraday Rotation} \label{polarization}

The polarimetric profiles of pulsars can be used to infer pulsar geometry \citep{1983ApJ...274..333R}. The MWA observations were recorded with full polarisation, however, to get the polarimetric profiles of pulsars and estimate the degree of linear polarisation, we first need to correct the effect of Faraday Rotation \citep[e.g.,][]{1975AJ.....80..794T}. We used the method described in \cite{2019PASA...36...25X}, which involves the application of rotation measure (RM) Synthesis technique to determine the rotation measure (RM, in units of rad m$^{-2}$) \citep{1966MNRAS.133...67B, 2005A&A...441.1217B, 2009A&A...503..409H}. We extracted the Stokes parameters $I$, $Q$, $U$, and $V$ as a function of frequency from the peak of the average total intensity pulse profile, using the pre-processing parts of the {\tt rmfit} \citep{2008MNRAS.386.1881N} function within the { \tt PSRCHIVE} package. These Stokes parameters were then used as input to the RM Synthesis code \footnote{\url{https://github.com/gheald/RMtoolkit}}. After correcting for the ionosphere contribution, we found the RM value of 69.5 $\pm$ 0.2 rad m$^{-2}$. \cite{10.1111/j.1365-2966.2007.12352.x} noted a slightly lower RM of 67.5 rad m$^{-2}$ for this pulsar using multiple frequencies from 0.69 to 3.1 GHz. The difference between the two values can be due to the change in the ionosphere and the local pulsar environment, considering the pulsar's proximity to the Galactic centre. The difference in RM values can also be expected given the time difference between observations and the pulsar space velocity.

%% file: Discussion.tex
\section{Discussion} \label{sec:discussion}

In this work, we have studied the pulse-to-pulse variability in single pulses using simultaneous wide-band observations made using uGMRT and MWA. The single pulses observed at multi-frequency uGMRT observations have an S/N of close to 100, making this pulsar a perfect source to study the wide-band single pulse behaviour. We employed a novel approach to calibrate the single pulses, using the flux densities estimated from the simultaneously recorded visibilities. In the following subsections, we discuss some of the results and implications of our studies. 


\subsection{Pulse Fluence Distribution}

Single-pulse fluence distributions can be used as tools to test models of the pulsar emission mechanism. Most models generally predict these distributions to be Gaussian, log-normal, or power-law \citep{1995MNRAS.276..372L,1998ApJ...506..341W}. For our case, a log-normal distribution well describes the fluence distribution of PSR J1820--0427 over a wide range of observed frequencies. 

Theories for wave growth in inhomogeneous media can be compared with pulsar observations. They can also be used to constrain the emission mechanisms and the potential source of these emissions. The Stochastic Growth Theory (SGT) \citep{RevModPhys.69.507, robinson1992clumpy, 1995PhPl....2.1466R, 1993ApJ...407..790R} describes self-consistently interacting systems where the interactions occur in an independent homogeneous medium and introduce the relevant distance and time scales. This theory predicts a log-normal distribution of the observed single pulse energies. 

\cite{2004MNRAS.353..270C} studied the phase-resolved single-pulse properties of slow-spinning pulsars PSR B0950+08 and PSR B1641--45, along with the Vela pulsar \citep{2003ASPC..302..191C, 2003MNRAS.343..512C, 2003MNRAS.343..523C}. Their investigation found that the pulsars showed log-normal fluence distributions over most of the pulse phase. They further reasoned that stochastic growth plays a pivotal role in producing pulsar emission, in which linear instabilities in inhomogeneous plasma generate the radio emission. The SGT also predicts a log-normal pulse fluence distribution seen in a broad population of pulsars, directly indicative of a changing SGT state \citep{2022ApJ...929...71W}. By interpreting the field statistics and the observed variability in a pure SGT, the associated pulsar emission mechanism is suggested to involve only linear processes. This implies that a plasma instability in an SGT state either directly generates the radiation, or else, it generates non-escaping waves that are transformed into escaping radiation by linear processes. 

\cite{2012MNRAS.423.1351B} studied the pulse fluence distributions of 315 pulsars at 1.4 GHz, and found that a log-normal distribution adequately describes the fluence distribution for a majority of pulsars. They performed fluence distribution tests to assess the shape of the probability distribution function. In their study, PSR J1820--0427 was fit by a log-normal distribution. Our analysis makes use of a much larger sample of single pulses, simultaneously observed at $<$1\,GHz, a much wider frequency range and examines the frequency evolution of the distribution. Our results therefore, extend their analysis with a log-normal pulse fluence distribution at a simultaneously observed wide frequency range from $\sim$300 to 750 MHz. A direct consequence of the combined result from our study and \cite{2012MNRAS.423.1351B} is that the primary pulsar emission mechanism is similar within the observed $\sim$1.5 GHz range. 

Our analysis reveals a small number of large amplitude pulses. However, they cannot be considered as "giant pulses", which are typically $\sim$2-3 orders of magnitude brighter than regular pulses \citep{2004ApJ...612..375C}. The arrival of these strong pulses had no phase dependence. This may suggest that the emission mechanism for the strong pulses is not any different from those responsible for the emission of  the regular (moderately bright) pulses.

\subsection{Spectral Analysis}\label{sec:sec5.2}

\subsubsection{Mean Flux Density Spectrum}

In general, pulsars are known to have a steep flux-density spectrum \citep{1973A&A....28..237S} at frequencies above $\sim$100 MHz. The study by \cite{1995MNRAS.273..411L} published spectra of 280 pulsars from measurements at multiple frequencies between 0.4 and 1.6 GHz. The resulting analysis showed that the average spectral index for the pulsar sample was $\alpha = -1.6 \pm 0.3$. Another study \citep{2000A&AS..147..195M} which extended the observed frequency range to 5 GHz, derived a mean value of $\alpha = -1.8 \pm 0.2$.

Some of the published literature \citep{2016MNRAS.461..908B, 2019ApJ...874...64Z} use a single power-law for the spectral behaviour of PSR J1820--0427. However, \cite{2017PASA...34...20M} suggested a broken power law to fit the pulsar spectra. The recent study by \cite{2022PASA...39...42L}, which spanned a similar frequency range,  also suggested a broken power law with the break frequency at 200 MHz, where below the break frequency $\alpha_{low} = 1.73 \pm 0.03$ and above 200 MHz $\alpha_{high} = -2.21 \pm 0.02$. However, in this study, we have combined 17 flux density data points, obtained from measurements made with MWA (185 MHz) and uGMRT (300-750 MHz) at multiple epochs, with several others from the published literature. With the addition of new data, flux density spectrum of PSR J1820--0427 now shows a low-frequency power-law turn-over (Fig. \ref{figure:spind}), much in contrast with previous results. The turn-over frequency with this model is 169 $\pm$ 5 MHz and the spectral index is -2.34 $\pm$ 0.01. The spectral index value obtained from our study is close to the value obtained in \cite{2022PASA...39...42L}, though only at high frequencies. Such low-frequency flux density measurements are crucial for constraining the spectral characteristics of pulsar emission at frequency ranges where the Square Kilometre Array (SKA-low) will be operational.


\subsubsection{Pulse-to-pulse Variability}

Single pulses from pulsars exhibit variabilities on multiple different time scales, and by examining these single pulses, we can constrain the pulsar emission mechanism. The intermittent behaviour of pulsars and the possibility of undetected populations raise questions about the presence of intermittent activity or emission variability in all pulsars and its impact on their evolution.For example, if the pulsar emission mechanism produces random intensity pulses, then one can expect a Gaussian distribution of pulse energies. However, any deviation from a Gaussian will point towards a non-random generation of intensities and hint toward a more complicated mechanism. Investigations of the individual pulses from slow pulsars revealed that their emission is quite erratic, especially compared to the stable average pulse profiles. Studying these variabilities is imperative in understanding the pulsar emission behaviour and propagation effects of the intervening plasma. In a recent study, \cite{2022A&A...661A.130A} performed a 2-dimensional particle-in-cell simulation to investigate the plasmoid formation due to magnetic reconnection in the pulsar magnetosphere as a possible source of pulse-to-pulse variability. According to their study, the origin of strong subpulses is related to plasmoid formation in the magnetosphere. These plasmoids produce bright, short subpulses, leading to a strong pulse-to-pulse variability. \cite{2022A&A...661A.130A} also provide some predictions about the flux density distribution of single pulses which can be applicable to single pulse studies. This kind of model, where bright subpulses are emitted due to the incoherent synchrotron radiation emitted in the pulsar wind current layer, can explain the high variability seen in pulsars like PSR J1820--0427.

\begin{figure}
    \centering
    \includegraphics[width=\columnwidth]{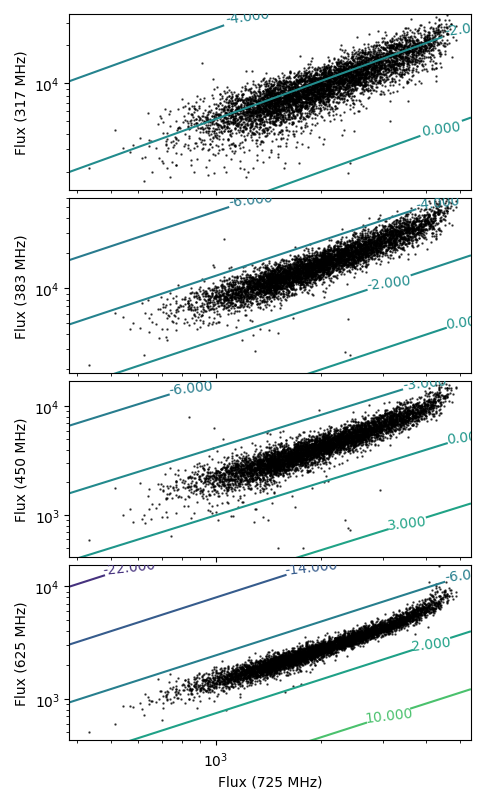}
    \caption{Flux vs flux plot between 725 MHz (x-axis) and other frequencies (y-axis). Contours show the lines of constant spectral index. The plot shows a tighter correlation between close frequency pairs. There also appears to be an upward trend for high-intensity pulses indicating steeper spectral index and a possibly different emission mechanism.}
    \label{fig:SI_contour}
\end{figure}

Utilising the advantage of the wide observing bandwidth, we inspected the correlation between the pulse intensities at all observed frequencies with the highest observed frequency (725 MHz) using the calibrated single pulse flux densities. Fig. \ref{fig:SI_contour} shows scatter plots between observed flux at particular frequencies vs the reference flux at 725 MHz. The contours show lines of constant spectral index. Spread in the spectral index values implies that the physical process giving rise to single pulses is chaotic, despite showing a global trend. We also found a peculiar behaviour of pulses at high flux densities, different from normal pulses. There appears to be an upward trend in all the subplots of Fig. \ref{fig:SI_contour}, indicating a steeper spectrum for the high-intensity pulses with respect to higher frequencies. Current literature lacks reference to any such observed behaviour in J1820–0427 or any other pulsars.

Thus, an emission process which generates pulses of variable intensity, along with a mechanism which increases coherence with pulse intensity and frequency, may explain the erratic single pulse emission behaviour of PSR J1820--0427. Increased coherence can also imply a suitable change in the local environment. Since we observe the steepening at higher frequencies (see Fig. \ref{fig:SI_contour}), implying lower emission heights (Radius to Frequency Mapping, \cite{ 1978ApJ...222.1006C}), the coherence in brighter pulses could be due to the closer proximity of the emission region to the pulsar surface, where the magnetospheric conditions are generally different. While this category of pulses may likely make up the tail end of the log-normal pulse fluence distribution that is observed, our observations seem to suggest an additional, intermediate category of pulses (somewhere between ``normal'' and ``giant'' pulses), which conform to large-amplitude single pulses seen in our data for PSR J1820–0427.

%% file: Conclusion.tex
\section{Summary} \label{sec:conclusion}

We have performed the first simultaneous multi-frequency single pulse analysis of PSR J1820$-$0427. In this work, we have used a high-quality data set from uGMRT at Band 3 (300-500 MHz) and Band 4 (550-750 MHz) and from MWA at 170-200 MHz. We estimated the flux densities using simultaneously recorded visibility data and used them to calibrate the single pulse data. Studies of single pulses can be insightful in understanding the pulsar emission process. The main findings from our analysis can be summarised below -
\begin{enumerate}
    \item We devised a novel method to calibrate the single pulse beamformed data using simultaneously recorded visibility data. This calibration method was applied to the pulsar data from uGMRT at Band 3 (300-500 MHz) and Band 4 (550-750 MHz) obtained in the phased array mode.
    \item We find that the single pulse fluences are best described using a log-normal distribution over a wide range of observed frequencies from 300 to 750 MHz. We further interpret our findings in terms of the Stochastic Growth Theory (SGT). Our results (at $\sim$0.1-1 GHz) are consistent with that from \cite{2012MNRAS.423.1351B} (at $\sim$1-2 GHz).
    \item  Using a rounded Pulse Broadening Function (PBF) corresponding to a thick scattering screen, for the measured pulse broadening time, we calculated a frequency scaling index, $ \beta \sim 3.5 \pm 0.3$. The reconstructed profile at the low-frequency MWA band revealed a post-cursor type feature in the emission, for which no clear counterparts are seen at higher frequencies (300-750 MHz). Since this feature is absent at the higher frequencies, it may either have a much steeper spectrum than the main profile, or alternatively, it is a feature that is visible at only lower frequencies (due to the specific beam/emission geometry of the pulsar).
    \item Using our multi-frequency observations at a large frequency band (170-750 MHz), in conjunction with published literature, we find that the mean flux density spectrum of PSR J1820--0427 is best described using a power-law function with a low-frequency turnover, in contradiction to the previously thought broken power-law. With this model, the turn-over frequency is 169 $\pm$ 5 MHz and the spectral index is -2.34 $\pm$ 0.01.
     \item We studied the pulse-to-pulse spectral index variation for PSR J1820--0427. The large scatter in the single pulse spectral index indicates a physical process less organised at short time scales.
     \item Our analysis also indicates a steeper spectrum for high-intensity pulses, as compared to the regular pulses. This anomalous behaviour may point towards an emission mechanism where the coherence increases with intensity and frequency.
\end{enumerate}

Overall, our work signifies the importance of high-sensitivity single pulse analysis over a wide frequency range to understand the intricacies of pulsar emission physics. Pulsars like J1820-0427, with their featureless profile, and no evidence of sub-pulse modulation such as nulling and drifting, are promising targets for detailed investigation of pulse fluence distributions and their frequency dependence. This is one of the less explored aspects of pulse-to-pulse variability, which can now be further investigated, thanks to the advent of uGMRT and its wide-band instrumentation. Other suitable targets include PSRs J0630--2834 and J1752--2806 which may allow extending the analysis down to the low frequencies of the MWA. 

Our work also underscores the importance of low-frequency measurements from facilities such as the MWA, e.g., for refining the spectral behaviour of pulsar emission. The emergence of an intriguing post-cursor type feature at  low frequencies illustrates the utility of de-scattering techniques that does not involve assumption of an intrinsic pulse shape. Analysis of this kind may allow us to explore the frequency evolution of average pulsar emission for pulsars where the profiles are substantially temporally broadened at  $\lesssim$ 300 MHz. 


